\documentstyle[sprocl]{article}

\begin{document}
\begin{flushright} DAMTP-96-103 \\ hep-th/9611135
\end{flushright}
\vspace{30pt}
\title{D-branes in the light-cone gauge and broken symmetries}

\author{ Michael Gutperle\footnote{M.Gutperle@damtp.cam.ac.uk} }

\address{DAMTP, Silver Street,\\Cambridge CB3 9EW, UK.}

\maketitle
\abstracts{
Boundary states for D-branes are constructed using the light cone
gauge. The D-brane breaks half the spacetime supersymmetry giving rise
to  fermionic zero modes living on the brane. The nonlinear
realization of the broken supersymmetry on the open string degrees of
freedom is analysed and the influence of boundary terms coming from
closed string vertex operators is  discussed.\\ 
(Contribution to the
workshop "Gauge Theories, Applied Supersymmetry and Quantum Gravity",
 London Imperial College, July 1996) }

\section{Boundary states and broken symmetries}
The introduction of boundaries breaks symmetries of conformal field
theories  because the
boundary relates left-moving and right-moving degrees of freedom. The
condition that the conformal invariance on the
half plane $H=\{z|Im(z)\leq 0\}$ is not completely broken
implies the continuity of  the holomorphic and antiholomorphic parts of
the stress energy tensor $T(z)=\bar{T}(\bar{z})$ for $Im(z)=0$. More
 generally  for a
current $J,\bar{J}$ of conformal weight $(h,h)$ the condition that the
 boundary
preserves the symmetry is given by the continuity
$J(z)=\bar{J}(\bar{z})$ for $z=\bar{z}$. Mapping the half plane into
the semi infinite cylinder via $z=\exp(\tau+i\sigma)$  we can
translate this
 condition into a
condition on a boundary state.
\begin{equation}
(J_n+(-1)^h\bar{J}_{-n})\mid B\rangle=0
\label{eq:current}
\end{equation}
We can define two currents acting on the boundary,
$J^{\pm}(\sigma)=J(z)\pm(-1)^h\bar J(\bar{z})$ where $J^-$ is the
current broken by the boundary conditions. Conformal boundary
conditions and boundary states are in one to one correspondence
\cite{cardy}.
 In the case of toroidally
compactified bosonic string theory with an  enhanced gauge symmetry group
$G\times G$ it was
shown in \cite{greengutperleb} that the scalars $J^{-a}$ correspond to
 Goldstone bosons living in the coset
$G\times G/G$ which is  defined trough  the breaking of the $G\times
G$
 closed string
symmetry to $G\times G/G$  by  the boundary.

D-branes \cite{Polchinski}  have been identified with stringy solitons carrying
RR-charges. A D-brane is a $p+1$ dimensional hyperplane
where open strings can end. The boundary conditions imposed on the
string coordinates are $\partial_n X^\alpha=0$ for $\alpha=0,\cdots,p$
and $\partial_t X^i=0$ for $i=p+1,\cdots,9$. The boundary conditions
for the world sheet fermions follows from (\ref{eq:current}) for the
super stresstensor $T_F=\psi_\mu\partial x^\mu$. The space time
 supersymmetry of D-branes has been
analysed using the R-NS formalism in \cite{Li,dealwis}. 
\section{Light cone supersymmetry }
The light cone Green-Schwarz formulation of the superstring has
manifest space time supersymmetry. We use a boundary state formulation
introduced in \cite{greengutperlec}    
where 
the string propagates on a cylinder and terminates at the proper time
$\tau=0$. The light cone gauge is imposed by  $X^+ = x^+ +
p^+ \tau$ and $X^+$ as well as $X^-$ satisfy Dirichlet boundary
conditions. The  coordinates transverse to the $\pm$ directions 
 satisfy
the boundary conditions
\begin{equation}\label{bosboun}
(\partial X^I - M^{I}_{\ \ J} \bar \partial X^J) |B\rangle
=0\end{equation}
where $M_{IJ}$ is an element of  $SO(8)$  and can be  written as,
\begin{equation}\label{eq:mgeneral}
M_{IJ} =   \exp\left\{\Omega_{KL}  \Sigma^{KL}_{IJ} \right\}
\end{equation}
where   $\Sigma^{KL}_{IJ} =  (\delta^K_{\ I}\delta^L_{\ J} - \delta^L_{\
I}\delta^K_{\ J}) $ are generators of $SO(8)$ transformations in
the vector
representation.
In this talk we are concerned with D-branes without boundary
condensates where 
the Neumann directions
are $\alpha = I =1, \cdots, p+1$ and  the Dirichlet
directions are  $i = I = p+2, \cdots ,8$
$M_{IJ}$  can be
written in block diagonal form,
\begin{equation}\label{eq:mdeff}
M_{IJ} = \pmatrix {- I_{p+1}& 0 \cr
      0 &   I_{7-p}  \cr}
\end{equation}
Each of the  sixteen-component supercharges of the type II theories
decompose
in the light-cone gauge  into two inequivalent  $SO(8)$ spinors
defined in
terms of the world-sheet fields $X^I$  and $S^a$ ($a=1,\cdots,8$) by,
\begin{equation}\label{susyalg1}
Q^a ={1\over \sqrt{2p^+}}\int_0^{\pi}d\sigma S^a(\sigma), \qquad
 {Q}^{\dot a}=  {1\over \pi \sqrt{p^+}}\int_0^{\pi }d\sigma
\gamma^I_{\dot a b}
 \partial X^I
 {S}^b(\sigma)
\end{equation}
for the left-moving charges and similarly  for the right-moving
charges, $\tilde Q^a$ and $\tilde Q^{\dot a}$ expressed in terms of the
right-moving coordinates. We define the combinations of the supercharges
\begin{equation}\label{eq:susyi}
Q^{\pm a} = (Q^a \pm iM_{ab} \tilde
Q^b)\;,\;
Q^{\pm\dot a}=  (Q^{\dot a} \pm i M_{\dot
a \dot b}
\tilde Q^{\dot b})
\end{equation}
The D-brane is a BPS-configuration which preserves half the spacetime
supersymmetry which means that 16 of the 32 supercharges are annihilated
by the boundary state.
\begin{equation}\label{eq:susyii}
  Q^{+ a} \mid B\rangle=0\;,\; Q^{+\dot a}\mid B\rangle=0
\end{equation}
These two conditions and the consistency with the light cone $N=2$
supersymmetry algebra are solved by $SO(8)$ rotations acting on the
spinors,
\begin{equation}\label{eq:mdef}
M_{ab} = \exp\{{1\over 2} \Omega_{IJ}\gamma^{IJ}_{ab}\}, \qquad \qquad
M_{\dot a\dot
b} = \exp\{{1\over 2}\Omega_{IJ}\gamma^{IJ}_{\dot a \dot b}\}
\end{equation}
where $\Omega_{IJ}$ is the same antisymmetric matrix (\ref{eq:mgeneral})  that
defined the $SO(8)$
rotation in the vector basis and $\gamma^{IJ} = 1/2(\gamma^I\gamma^J -
\gamma^J\gamma^I)$. For the boundary condition given in (\ref{eq:mdeff}) the
solution is simply $M_{ab}=\gamma^1\cdots\gamma^{p+1}$.

The boundary state that solves (\ref{eq:susyii}) can be obtained
\begin{eqnarray}\label{eq:boundarystate2}
|  B\rangle & =\exp\sum_{n>0}\left(   {1\over n}  M_{IJ}
\alpha^I_{-n}\tilde{\alpha}^J_{-n} -
i  M_{ab}S^a_{-n}\tilde{S}^b_{-n}\right)|B_0\rangle,\nonumber \\
&= R(M) \exp \sum_{n>0} \left({1\over n}
\alpha^I_{-n}\tilde{\alpha}^I_{-n}  -
i   S^a_{-n}\tilde{S}^a_{-n}\right)|B_0\rangle
\end{eqnarray}
where the zero-mode factor is given by
\begin{equation}\label{eq:boundarystate1}
|  B_0\rangle =C\left(   M_{IJ} |  I\rangle | J\rangle +i
M_{\dot{a}\dot{b}} |
\dot{a}\rangle| \dot{b}\rangle \right)
\end{equation}
The normalisation factor $C$ is determined by comparing the open
string one loop
partition function with the closed string vacuum to vacuum transition
calculated using the boundary state (\ref{eq:boundarystate2}).
The  fields of IIB supergravity expressed as a light-cone superfield by
introducing Grassmann coordinates \cite{brink}
$\theta^a=S^a_0-i\bar{S}^a_0$ and $p^+\partial/\partial
\theta^a=S^a_0+i\bar{S}^a_0$ which satisfy
$\{\partial/\partial\theta^a,\theta^b\}=\delta^{ab}$. To incorporate the
D-brane boundary conditions modified Grassman coordinates are
introduced by 
\begin{equation}\label{eq:newthet}
\hat \theta = {1\over 2}(1 + M)_{ab}\theta^b + {p^+ \over 2} (1-
M)_{ab}{\partial\over
\partial \theta^b}
\end{equation}
and its conjugate
\begin{equation}\label{eq:newpart}
{\partial \over \partial \hat \theta^a} = {1\over 2p^+}  (1-M)_{ab}
\theta^b + {1\over 2}(1
+ M)_{ab}{\partial \over \partial \theta^b} .
\end{equation}The conditions (\ref{eq:susyii}) imply  $
{\partial}/{\partial\hat \theta^a}| B_0\rangle=0$ which means  that the
D-brane can be interpreted as a bottom part of a IIB multiplet. Note
that the higher terms in the multiplet are obtained by multiplying
with up to eight $\hat{\theta}$ which corresponds to the emission of a zero
momentum fermions. The higher terms in the multiplet are important when
one considers scattering of D-branes with helicity flip which
constitutes a change in the D-brane state during the scattering process.
\section{Nonlinear realization of supersymmetry}
The massless  bosonic and fermionic open-string vertex operator  are 
given by
\begin{equation}\label{eq:bosvert}
V_B(\zeta,k) = (\zeta^IB^I-\zeta^-)e^{ikX}\;,\;V_F(u,k) = 
 (u^aF^a+u^{\dot{a}}F^{\dot{a}})e^{ikX},
\end{equation}
where
\begin{eqnarray}\label{eq:bosfact}
B^{I}  &=& \partial
 X^I-\frac{1}{2}S^{a}(z)\gamma^{IJ}_{ab}S^b(z)k^J\;,\;
 F^a = S^a(z)\\
F^{\dot{a}} &=&  \gamma^I_{a\dot{a}}S^{a}(z)\partial
X^I+\frac{1}{6}:
\gamma^I_{a\dot{a}}S^a(z)S^b(z)\gamma^{IJ}_{bc}S^c(z):k^J
\end{eqnarray}
The   16 components of the supersymmetry act on the
vertex operators in the following way,
\begin{eqnarray}\label{eq:transvert}
\delta_\eta V_B=  [\eta^aQ^a,V_B(\zeta)]&=&V_F(\tilde{u}) \nonumber \\
\delta_\eta V_F= [\eta^aQ^a,V_F(u)]&=&V_B(\tilde{\zeta}) \nonumber \\
\delta_\epsilon V_B=
[\epsilon^{\dot{a}}Q^{\dot{a}},V_B(\zeta)]&=&V_F(\tilde{\tilde{u}})+
\epsilon^{\dot{a}}\partial_z W^{\dot{a}}_B(\zeta,k,z) \nonumber\\
\delta_\epsilon V_F=
[\epsilon^{\dot{a}}Q^{\dot{a}},V_F(u)]&=&V_B(\tilde{\tilde{\zeta}})+
\epsilon^{\dot{a}}\partial_z W^{\dot{a}}_F(u,k,z)
\end{eqnarray}
Supersymmetry transformations on the $SO(8)$ components of the massless
open-string fields take the form,
\begin{eqnarray}\label{eq:susytr}
\tilde{\zeta}^I&=&  \eta^a\gamma^I_{a\dot{a}}u^{\dot{a}} \sim 0 ,\qquad
\tilde{\tilde{\zeta}}^I = \sqrt{\frac{1}{2}} \epsilon^{\dot{a}}
\gamma^I_{a\dot{a}}u^a + \frac{\sqrt{2}} {k^+}
\epsilon^{\dot{a}}u^{\dot{a}}k^I
=
k^I \epsilon^{\dot a} \gamma^{IJ}_{\dot a \dot b} v^{\dot b} ,
\nonumber \\
 \tilde{v}^{\dot{a}} &=&   \eta^a\gamma^I_{a\dot{a}}\zeta^I , \qquad
\tilde{\tilde{v}}^{\dot{a}}={1\over k^+} \sqrt{\frac{1}{2}} \left(
\epsilon^{\dot{a}}\gamma^{IJ}_{\dot{a} \dot{b}}k^I\zeta^J+
\epsilon^{\dot{a}}
\zeta^Ik^I\right)
\end{eqnarray}
The total derivative terms in (\ref{eq:transvert}) are given by 
\cite{greenseiberg}
\begin{eqnarray}\label{eq:totalder1}
W_B^{\dot{a}}&=&\sqrt{2}\gamma^{I}_{a\dot{a}}\zeta^IS^ae^{ikX}\nonumber\\
W_F^{\dot{a}}&=&\sqrt{2}v^{\dot{a}}e^{ikX} + \frac{\sqrt{2}} {8}
(\gamma^{IJ}u)^{\dot{a}} S^b \gamma^{IJ}_{bc} S^ce^{ikX}\label{eq:totaldB}.
\end{eqnarray}
 The
amplitude with $n$
bosonic open-string ground states has the form,
\begin{equation}\label{eq:openamplitude}
A_n(\Psi\mid \zeta_1,\cdots,\zeta_n)=\int
d\sigma_1..d\sigma_n\langle
\Psi\mid
V_B(\zeta_1,k_1,z_1)\cdots
V_B(\zeta_n,k_n,z_n)\mid B\rangle.
\end{equation}
A supersymmetry transformation of  this amplitude  is obtained  by
substituting
a transformed wavefunction $\tilde{\zeta_1}$ or
$\tilde{\tilde{\zeta_1}} $into
$A_n$. Using (\ref{eq:transvert}) the transformed vertex operator can
be written
as a
commutator of a supersymmetry generator and a fermionic vertex
operator.  Thus, for  the linearly realized components of the conserved
supersymmetry the vertex,
\begin{equation}\label{eq:commutat}
V_B(\tilde{\zeta})=\eta^a_+ Q^{+ a}V_F(u)-V_F(u)\eta^a_+  Q^{+ a},
\end{equation}
can be inserted into (\ref{eq:openamplitude}) and  the $Q^+$ in the
first term
acts to the left  on the closed string state $\Psi$ giving a
transformed state,
$\delta_{\eta_{\pm}}\Psi$.   The $Q^+$ in the second term is moved to the
right and gives transformed open string vertex operators until it hits
the boundary where it is annihilated  by the boundary state.   Thus the
conserved
 supersymmetry relates S-matrix elements with  $n+1$ bosonic states
(including
the one (bosonic) closed-string end-state)  to elements with $n-1$
bosonic and
2 fermionic states,
\begin{eqnarray}\label{eq:nongold}
& A_n(\Psi\mid\tilde{\zeta}_1,\zeta_2,\cdots,\zeta_n)
=A_n(\delta_{\eta^+}
\Psi\mid u_1,\zeta_2,\cdots,\zeta_n) + A_n(\Psi\mid
u_1,\tilde{u}_2,\zeta_3\cdots,\zeta_n)\nonumber\\
&+ \cdots\nonumber +  A_n(\Psi\mid
u_1,\zeta_2,\cdots,\zeta_{n-1},\tilde{u}_n).
\end{eqnarray}
 This corresponds to the linearly realized
supersymmetry which is not broken by the boundary state.   A
similar analysis
applies to the non-linearly realized conserved supercharge,
$Q^{+\dot a}$ with
$\tilde \zeta$ and $\tilde u$ replaced by $\tilde{\tilde \zeta}$ and
$\tilde{\tilde u}$.

The supercharge $Q^{-a }$ is not annihilated by the boundary so
that  similar
manipulations for these supercharges leave a residual term.  This  is
proportional to  $\eta^a_- Q^{-a}|B\rangle$, which has the form of the
fermion emission vertex   (\ref{eq:bosvert}) acting on the boundary, in which
the
supersymmetry parameter $\eta_-^a$ is the  wave function.
Therefore,  the
amplitude with $n+1$ bosonic states is related by the $Q^-$
supersymmetry  to a
sum of terms with $n-1$ bosons and two fermions, together with an
extra term
which has  an extra zero-momentum fermion insertion -- it has a
total of $n$
bosons and two fermions. This term can also be interpreted as a shift
in the fermionic collective coordinate of the boundary $\delta 
\theta^a=\eta^{-a}$.
\begin{eqnarray}\label{eq:goldstinoo}
& A_n(\Psi |  {\tilde{\zeta}}_1,\zeta_2,\cdots,\zeta_n)=
A_n(\delta_{\eta^-}
\Psi | u_1,\zeta_2,\cdots,\zeta_n) + A_n(\Psi  | u_1,
{\tilde{u}}_2,\zeta_3\cdots,\zeta_n)\nonumber\\
&+\cdots +  A_n(\Psi |  u_1,\zeta_2,\cdots,\zeta_{n-1}, {\tilde{u}}_n) +
A_{n+1}(\Psi\mid u_1,\zeta_2,\cdots,\zeta_n,\eta^{ - })
\end{eqnarray}
This  is the S-matrix statement of the nonlinear realization of  the
spontaneously broken $Q^-$ supersymmetry \cite{greengutperleb}.
 The
corresponding
analysis with the non-linearly realized supercharge $Q^{- \dot a}$
leads to the
same relationship between amplitudes but with $\tilde \zeta$ and
$\tilde u$
replaced with $\tilde{\tilde \zeta}$ and $\tilde {\tilde u}$.
Higher-order terms give rise to S-matrix elements with arbitrary
numbers of soft fermions. 

\section{Closed string symmetries and collective coordinates}
The vertex operator for massless NS-NS tensors is given by a product
of two open string vertices (\ref{eq:bosvert}).
Consider gauge transformations on the graviton 
$\delta\zeta_{IJ}=k_{(I}\Lambda_{J)}$ under which the  corresponding
vertex operator becomes  a total
derivative on the world sheet. 
\begin{eqnarray}
\delta\zeta_{IJ}\int d^2z V^I(z)\bar{V}^J(\bar{z})\mid B\rangle &=& 
\Lambda_I(\delta^{IJ}+M^{IJ})\oint d\sigma \partial X^J\mid
B\rangle\nonumber\\
&+&\oint d\sigma S^a (\gamma^{IK}-M^t\gamma^{IK}M)_{ab}S^bk_K\mid 
B\rangle\nonumber
\end{eqnarray}
The boundary term  induces a shift of the
position of the D-brane in the transverse directions $Y^I\to
Y^I+\Lambda^I, I=p+2,\cdots,7-p$. The second term vanishes when one
considers the limit $k^I\to 0$.  

The undotted components of the two gravitini also produce  boundary
terms upon gauge transformation $\delta\zeta_{Ia}=u^ak^I$
\begin{eqnarray}
\delta\zeta_{Ia} \int d^2z V_B^I(z)\bar{V}_F^a(\bar{z})&=& u^a\oint
d\sigma 
S^a\mid  B\rangle\nonumber\\
\delta\hat{\zeta}_{Ia}\int d^2zV_F^a(z)\bar{V}_B^I(\bar{z})&=&
 u^a\oint d\sigma  \bar{S}^a\mid  B\rangle
\end{eqnarray}
The gauge transformation of the combination  $\zeta_{Ia}+
iM_{ab}\hat{\zeta}_{Ib}$ induces a shift of the
fermionic collective coordinate $\delta\hat{\theta}=u^a$ given by the 
insertion of a zero
momentum fermionic vertex operator at the boundary. The other
gravitino given by $\zeta_{Ia}-iM_{ab}\hat{\zeta}_{Ib}$ does not give
rise to a boundary term and correspond to the unbroken
supersymmetry. Similarly the combiantion
$\zeta_{I\dot{a}}+iM_{\dot{a}\dot{b}}\hat{\zeta}_{I\dot{b}}$ produces a
boundary term proportiaonl to $Q^{\dot{a}-}$.

Applying the broken supersymmetry to the boundary states corresponds
to inserting fermionic zero modes on the D-brane. The bosonic zero
modes correspond to the broken translational and internal symmetries
on the brane and should be generated by acting with the unbroken
supersymmetry on the fermionic zero mode.
\begin{equation}
u_1^{\dot{a}}Q^{\dot{a}+}u_2^a Q^{b-}\mid B\rangle=
u_1^{\dot{a}}\gamma^I_{\dot{a}b}u_2^b
\oint (\partial X^I+M^{IJ}\bar{\partial} X^J)\mid B\rangle
\end{equation}
The unbroken supersymmetry acting on a fermonic zero mode
creates a vertex which is proportional to the momentum of the boundary
state in the Dirichlet directions since $p^I\mid
B(y)\rangle=i\partial/\partial y^I \mid
B(y)\rangle$ this term creates a shift in the
transverse position of the D-brane $\delta Y^I=u_1\gamma^Iu_2$. In the
Neumann directions the operator is proportional to the winding number
of the boundary and it does not contribute for flat D-branes.
\section{Contact terms}
The fact that total derivative terms appear in the representation of
SUSY on the massless vertex operators becomes important when one
considers the supersymmetry in the presence of closed string vertex
operators. For simplicity consider  the D-instanton and the action of
the unbroken supersymmetry on the NS-NS antisymmetric tensor.
\begin{equation}\label{eq:susyvar1}
(Q^{\dot{a}}+i\bar{Q}^{\dot{a}})\xi_{IJ}\int d^2z V^I(z)\bar{V}^J(\bar{z})\mid
B\rangle
 =\xi_{IJ}\oint d\sigma (W^{I\;\dot{a}}\bar{V}^J+iV^I\bar{W}^{J\dot{a}})\mid
B\rangle
\end{equation}
inserting (\ref{eq:bosfact}) and (\ref{eq:totalder1}) into
 (\ref{eq:susyvar1}) then gives 
\begin{eqnarray}\label{eq:boundsusy1}
&&\xi_{IJ}\oint d\sigma
\left(\gamma^I_{a\dot{a}}S^a\partial_{\bar{z}}X^J+
i\gamma^J_{a\dot{a}}\bar{S}^a\partial_{{z}}X^I\right)\nonumber\\
&&-{1\over
  2}\xi_{IJ}\oint d\sigma
\left(\gamma^I_{a\dot{a}}S^ak^N\bar{S}^b\gamma^{IN}_{bc}\bar{S}^c
-ik^N{S}^b\gamma^{IN}_{bc}{S}^c\gamma^J_{a\dot{a}}\bar{S}^a\right)
\end{eqnarray}
The boundary conditions can be  used to
express the right-moving fields in terms of the left-moving ones.
It is  then easy to see that the second term in (\ref{eq:boundsusy1})
vanishes for an antisymmetric $\xi_{IJ}$, but the first term is
nonzero and corresponds to a violation of space time supersymmetry.
This violation  can be cancelled 
by adding a boundary term of the form
\begin{equation}
\xi_{IJ}\oint d\sigma S^a(\sigma)\gamma^{IJ}_{ab}S^b(\sigma)\mid
B\rangle=\xi_{IJ}R_0^{IJ}\mid B\rangle
\end{equation}
Note that the boundary term is a helicity operator which  rotates the
leftmoving  fermionic modes $S_n$
leaving the bosonic modes and the right-moving fermions  untouched.
The fact that the boundary condition on the bosonic coordinates 
$X$ are not changed means that the modified boundary state represents
still a D-instanton, but it couples to different closed string
fields.
The supersymmetry variation of this boundary term gives
\begin{eqnarray}
(Q^{a}+i\bar{Q}^{a})\xi_{IJ}R_0^{IJ}\mid B\rangle&=&
\xi_{IJ}\gamma^{IJ}_{ab}Q^a\mid B\rangle\\
(Q^{\dot{a}}+i\bar{Q}^{\dot{a}})\xi_{IJ}R_0^{IJ}\mid B\rangle
&=&\xi_{IJ}\gamma^{IJ}_{\dot{a}\dot{b}}Q^{\dot{b}}+\xi_{IJ}\oint d\sigma
\partial_z X^I\gamma^{J}_{\dot{a}a}S^a\mid B\rangle
\end{eqnarray} 
The presence of the boundary term hence induces a modification of the
conservation of the supersymmetry charge by the modified boundary
state and cancel the boundary term coming from the nonlinearly
realized supersymmetry.  For the massless part of the boundary state we have
\begin{equation}
\xi_{IJ}R^{IJ}(\mid N\rangle{\mid \bar{N}\rangle}+i\mid \dot{a}\rangle\mid
 \bar{ \dot{a}}\rangle)=\xi^{IJ}\mid I\rangle\mid \bar{J}>+
i\xi^{IJ}\gamma^{IJ}_{\dot{a}\dot{b}}\mid \dot{a}\rangle\mid
  \bar{\dot{a}}\rangle
\end{equation}
i.e. the boundary state couples to a specific combination of the NS-NS
and R-R antisymmetric tensor states.

It is also interesting to note that the bispinor states do not give a
contact  term. 
The supersymmetry transformation on a bispinor state is given by
\begin{equation}\label{RRbound}
\epsilon^{\dot{a}}(Q^{\dot{a}}+i\bar{Q}^{\dot{a}})u_a\bar{u}_b\int
 d^2z
 V^a(z)\bar{V}^b(\bar{z})\mid
B\rangle
 =\oint d\sigma \epsilon^{\dot{a}}(v^{\dot{a}}\bar{u}^{b}\bar{S}^b
-i\bar{v}^{\dot{a}}u^bS^b)e^{iky}
\mid B\rangle
\end{equation}
We can define the new susyparameter $\epsilon^{\prime
  a}=\epsilon^{\dot{a}}k_I\gamma^I_{a\dot{a}}$. Using the boundary
conditions (\ref{RRbound})  can be written in the following form
\begin{equation}
\epsilon^{\prime
  a}\oint (u^a\bar{u}^b-\bar{u}^au^b)S^a\mid B\rangle=\epsilon^{\prime
  a}\xi^{IJ}\gamma^{IJ}_{ab}Q^b\mid B\rangle
\end{equation}
Where the antisymmetric part of the bispinor is written in terms of
the corresponding AST wave function using $u_{[a}u_{b]}=
\xi_{IJ}\gamma^{IJ}_{ab}$

\section*{Acknowledgments}
I would like to thank M.B. Green for collaboration and discussion on
matters presented in this talk. I gratefully
acknowledge support by EPSRC and a Pannett Research Studentship of
Churchill College, Cambridge.

\section*{References}
\newcommand{\Journal}[4]{{#1} {\bf #2}, #3 (#4)}
\newcommand{\NCA}{\em Nuovo Cimento}
\newcommand{\NIM}{\em Nucl. Instrum. Methods}
\newcommand{\NIMA}{{\em Nucl. Instrum. Methods} A}
\newcommand{\NPB}{{\em Nucl. Phys.} B}
\newcommand{\PLB}{{\em Phys. Lett.}  B}
\newcommand{\PRL}{\em Phys. Rev. Lett.}
\newcommand{\PRD}{{\em Phys. Rev.} D}
\newcommand{\ZPC}{{\em Z. Phys.} C}


\begin{thebibliography}{99}
\bibitem{cardy} J. Cardy, \Journal{\NPB}{324}{581}{1989}
\bibitem{greengutperleb} M.B. Green and M. Gutperle,
  \Journal{\NPB}{460}{77}
{1996} 
\bibitem{Polchinski} J. Polchinski, S. Chaudhuri and  C.V. Johnson,
  {\bf hep-th/9602052} and references therein
\bibitem{Li}M. Li, \Journal{\NPB}{460}{351}{1996}
\bibitem{dealwis}S.P. de Alwis and K. Sato, \Journal{\PRD}{53}{7187}{1996}
\bibitem{greengutperlec} M.B. Green and M. Gutperle,
  \Journal{\NPB}{476}{484}{1996}
\bibitem{brink}M.B. Green, J.H. Schwarz and L. Brink,
  \Journal{\NPB}{219}
{437}{1983}
\bibitem{greenseiberg} M.B. Green and N. Seiberg,
  \Journal{\NPB}{299}{599}
{1987}
\end{thebibliography}
\end{document}